# Quantum error correction with silicon spin qubits

Kenta Takeda,[1, *] Akito Noiri,[1] Takashi Nakajima,[1] Takashi Kobayashi,[2] and Seigo Tarucha[1, 2 *]

[1]Center for Emergent Matter Science (CEMS), RIKEN, Wako-shi, Saitama, 351-0198, Japan
[2]Center for Quantum Computing (RQC), RIKEN, Wako-shi, Saitama 351-0198, Japan

[*]Correspondence to: Kenta Takeda (kenta.takeda@riken.jp) or Seigo Tarucha (tarucha@riken.jp)

Main text:

Large-scale quantum computers rely on quantum error correction to protect the fragile quantum information[1,2]. Among the possible candidates of quantum computing devices, silicon-based spin qubits hold a great promise due to their compatibility to mature nanofabrication technologies for scaling up[3]. Recent advances in silicon-based qubits have enabled the implementations of high quality one and two qubit systems[4–6]. However, the demonstration of quantum error correction, which requires three or more coupled qubits[1] and often involves a three-qubit gate[7–9], remains an open challenge. Here, we demonstrate a three-qubit phase correcting code in silicon, where an encoded three-qubit state is protected against any phase-flip error on one of the three qubits. The correction to this encoded state is performed by a three-qubit conditional rotation, which we implement by an efficient single-step resonantly driven iToffoli gate. As expected, the error correction mitigates the errors due to one qubit phase-flip as well as the intrinsic dephasing due to quasi-static phase noise. These results show a successful implementation of quantum error correction and the potential of silicon-based platform for large-scale quantum computing.

Quantum computing takes advantage of quantum superposition and entanglement to accelerate the computational tasks[10,11]. However, these quantum properties are sensitive to decoherence errors due to energy relaxation and dephasing. As the number of qubits increases and/or the computational tasks become more complex, the errors cause exponential reduction of the accuracy of computational results. Quantum error correction (QEC) is a protocol to circumvent this problem by distributing the quantum information across a larger multiqubit entangled state so that the errors can be detected and corrected[12]. Its basic concept has been demonstrated in various platforms such as nuclear magnetic resonance[7,13], trapped ions[8,14], nitrogen vacancy centers[15], and superconducting circuits[9,16,17] and has served as an important benchmark of the qubit systems. Silicon-based spin qubits have emerged as a qubit platform in the last decade, and there have been rapid progress in long coherence times[18,19], high-fidelity universal quantum gates[4–6], high-temperature operation[20,21], and generation of three-qubit entanglement[22].

In this paper, we demonstrate a three-qubit quantum error correcting code in silicon. The key elements of QEC are the abilities to encode, decode, and correct the quantum state. The minimum setup to demonstrate these elements is a three-qubit system (Fig. 1a) to correct one data qubit ($Q_2$) coupled with two ancilla qubits ($Q_1$ and $Q_3$). First, a three-qubit entangled state is created in the encoding (repetition code), and then the error on the encoded state is mapped to the ancilla measurement operator $Z_1 Z_3$ in the decoding. The correction to the data qubit can be performed by a rotation conditioned on $Z_1 Z_3$. This is achieved by a three-qubit iToffoli gate, which coherently rotates the data qubit conditioned on the ancilla spin polarization, hence avoiding the dynamic feedback control that is challenging with the present silicon qubits. With this approach, we demonstrate that one qubit phase-flip error can be corrected, and the intrinsic ensemble spin dephasing can be mitigated.

Our sample is a gate-defined triple quantum dot in an isotopically natural silicon/silicon-germanium (Si/SiGe) heterostructure. Three layers of overlapping aluminum gates[23] are used to control the triple-dot confinement. A micro-magnet is fabricated on top of the aluminum gates to provide a local magnetic field gradient[24]. As schematically shown in Fig. 1b, we configure the gate voltages so that only one electron is confined under each of the plunger gates (P1, P2, and P3), and the inter-dot tunnel coupling is controlled by the barrier gates (B2 and B3). Measurement of the triple-dot charge configuration is performed by monitoring the conductance of the nearby charge sensor quantum dot using the radio-frequency reflectometry technique[25,26]. An in-plane external magnetic field of $B_{\text{ext}} = 0.607$ T is applied using a superconducting magnet. We utilize the Zeeman-split spin-1/2 states of the single electrons as our spin qubits (labeled $Q_1$, $Q_2$, and $Q_3$ in Fig. 1b, c). The Zeeman energy splitting (~20 GHz) much larger than the thermal excitation energy (~0.8 GHz or ~40 mK) enables initialization and readout of the three-spin state by the combination of energy-selective tunneling[27], shuttling[28], and controlled-rotation (see Methods and Extended Data Fig. 1 for the full details of the sequence).

The single-qubit rotations are performed by applying resonant microwave pulses (see Methods and Extended Data Fig. 2). The microwave pulse displaces the quantum dot position, effectively creating an oscillating transverse magnetic field that induces electric-dipole spin resonance[24]. The two-qubit controlled phase (CZ) gate is implemented by adiabatically pulsing the exchange couplings $J_{12}$ and $J_{23}$ by the barrier gates B2 and B3, respectively (see Methods and Extended Data Fig. 3). To operate the qubit close to the charge-symmetry point, the capacitive crosstalk between the plunger and barrier gates is suppressed by the virtual gate technique (see Methods). The spin qubits herein have an average $T_1$ relaxation time of 22 ms, inhomogeneous dephasing time $T_2^*$ of 1.8 μs, and Hahn echo dephasing time $T_2^H$ of 43 μs (Extended Data Fig. 4). Since electron spins have orders of magnitude longer $T_1$ times compared to the dephasing times $T_2^*$ and $T_2^H$, we focus on the implementation of a phase-flip correction code in this work, while a bit-flip correction code can easily be assembled by introducing additional single-qubit rotations.

First, we demonstrate the ability to encode and decode the data qubit state. For simplicity, here we perform encoding of the most nontrivial cases where the input state is on the equator of the Bloch sphere, $Q_2 = (|\downarrow\rangle + e^{i\phi}|\uparrow\rangle)/\sqrt{2}$ (Fig. 2a, $\phi$ is an azimuthal angle). Encoding of this state results in a maximally entangled three-qubit Greenberger–Horne–Zeilinger (GHZ) state $|GHZ_\phi\rangle = (|\downarrow\downarrow\downarrow\rangle + e^{i\phi}|\uparrow\uparrow\uparrow\rangle)/\sqrt{2}$. The controlled-not (CNOT) gates used in the encoding are decomposed to native CZ gates combined with the decoupling pulses to mitigate the local quasi-static phase noise. For the QEC implementation, a crucial property is that the encoded state is a genuine three-qubit GHZ-class state. We confirm this by characterizing the generated state using three-qubit quantum state tomography (Methods). In Fig. 2b (2c), the real part of the measured experimental density matrix $\rho$ for $\phi = 0$ ($\pi$) is plotted. We evaluate the state fidelities $F = \langle GHZ_\phi|\rho|GHZ_\phi\rangle$ for various $\phi$ (Fig. 2d) and confirm that all the states have fidelities above 0.75, the threshold to witness genuine GHZ-class states.

For correcting the decoded state, we implement a Toffoli-class three-qubit gate. The standard three-qubit Toffoli gate can be synthesized from 12 CNOT and 2 single-qubit gates[29,30] (excluding T gates that can be implemented in software), albeit that decoherence in our device does not allow this implementation with a reasonable fidelity. Alternatively, we utilize a single-step, resonantly driven iToffoli gate implemented by a resonant $\pi$ pulse in the presence of simultaneous nearest neighbor exchange couplings (Fig. 2e). Without the exchange couplings (left side of Fig. 2e), the four transitions associated with the $Q_2$ rotation are degenerate with a resonance frequency of $f_0$. The finite exchange couplings shift down the energy levels of the anti-parallel spin configurations. As a result, the resonance frequency of $Q_2$ is modulated as $f_0 + s_1 J_{12} + s_3 J_{23}$, where $s_i = \pm 1/2$ is the spin number of $Q_i$. Under the condition with $J_{12} = J_{23}$ required for conditional phase synchronization (see Methods), a rotation of $Q_2$ with $Q_1 Q_3 = |\downarrow\downarrow\rangle$ or $|\uparrow\uparrow\rangle$ corresponds to a controlled-controlled-rotation.

Figure 2f shows the spectra of $Q_2$ with four different ancilla qubit states $Q_1 Q_3 = |\downarrow\downarrow\rangle$, $|\downarrow\uparrow\rangle$, $|\uparrow\downarrow\rangle$, and $|\uparrow\uparrow\rangle$ at $J_{12} = J_{23} = 4.5$ MHz, where we observe the peak positions as expected from the exchange couplings. We use a resonant $\pi$ pulse at $f_{MW} = f_1$ ($Q_1 Q_3 = |\downarrow\downarrow\rangle$) to implement our iToffoli gate since this transition yields the highest visibility[31]. The iToffoli gate is a three-qubit gate equivalent to a Toffoli gate with an additional phase factor of $i$ on the ancilla qubits. In order to characterize its property, we prepare the eight possible three-spin eigenstates, apply the iToffoli gate, and perform three-spin projective measurement (Fig. 2g, h). The readout errors are removed from the data based on the measured readout fidelities (see Methods). The Rabi frequency is chosen so that the off-resonant rotations for the $Q_1Q_3 = |\downarrow\uparrow\rangle/|\uparrow\downarrow\rangle$ subspaces are synchronized (see Methods). In Fig. 2h, as expected, the populations of $|\downarrow\downarrow\downarrow\rangle$ and $|\downarrow\uparrow\downarrow\rangle$ states are swapped, whereas the other states are essentially unaffected. From this result, we obtain a population transfer fidelity of our iToffoli gate as $\text{Tr}(U_\text{expt}U_\text{ideal})/8 = 0.96$, where $U_\text{expt}$ ($U_\text{ideal}$) represents the experimental (ideal) classical action. In addition, we perform a calibration of the pulse duration and timing to eliminate unwanted phase accumulation on $Q_2$ (see Methods). Note that the dephasing and phase accumulation on the ancilla qubits do not affect the error correction outcome.

We then turn to the implementation of the phase-flip correcting code. Figure 3a shows the quantum circuit diagram. The three-qubit operation U serves to encode the data qubit state $|\psi\rangle$ to the three-qubit entangled state. The exact implementation of U is shown in the bottom half of the figure, and it is equivalent to the two CNOT gates shown in Fig. 2a except for the single-qubit gates that do not affect the function of the QEC. Here, the data qubit state $|\psi\rangle = \alpha|\downarrow\rangle + \beta|\uparrow\rangle$ is encoded to a phase-sensitive three-qubit state $\alpha|+++\rangle + \beta|---\rangle$, where $|\pm\rangle = (|\downarrow\rangle \pm |\uparrow\rangle)/\sqrt{2}$ are the eigenstates of the Pauli X operator. For a phase-flip error with a flip rate of $p$ on $Q_2$, the decoded state is $\sqrt{1-p}|\downarrow\rangle(\alpha|\downarrow\rangle + \beta|\uparrow\rangle)|\downarrow\rangle + \sqrt{p}|\uparrow\rangle(\beta|\downarrow\rangle + \alpha|\uparrow\rangle)|\uparrow\rangle$ (see Extended Data Table 1 for the cases with an error on ancilla). The correcting procedure is implemented so that $Q_2$ is flipped only when $Q_1Q_3=|\uparrow\uparrow\rangle$ by applying $\pi$ pulses on the ancilla qubits followed by the iToffoli gate, resulting in a product state of $Q_2 = \alpha|\downarrow\rangle + \beta|\uparrow\rangle$ and $Q_1Q_3 = \sqrt{1-p}|\uparrow\uparrow\rangle + i\sqrt{p}|\downarrow\downarrow\rangle$. Now, the data qubit state is the same as the input state regardless of $p$. This is demonstrated in Fig. 3b, where we estimate the process fidelity of the data qubit for various one qubit errors (see Methods for the details of quantum process tomography). The one qubit error is a phase rotation with a known rotation angle $\theta$, which is equivalent to a phase-flip error with $p = \sin^2(\theta/2)$. Therefore, without the correction, the process fidelity oscillates as a function of $\theta$, shown as the black points. With the correction, the oscillation vanishes, and it confirms the basic function of the phase-flip correcting code. When there is no error ($\theta = 0$), the process fidelity slightly decreases after the correction. This can be attributed to the infidelity of the iToffoli gate projected to the data qubit subspace. Furthermore, we show that the state of ancilla qubits reflects the error on the encoded qubit state (error detection). We measure the joint probability of the ancilla qubits $Q_1$ and $Q_3$ for the four possible cases with no error or a full $\pi$ flip error. We observe that the measured ancilla states correctly reflect the error occurred to the encoded three-qubit state (Fig. 3c).

Errors in actual quantum computers likely occur on all qubits simultaneously rather than on only one of the qubits. We verify the performance of our error correcting code in such a case where all errors have the same effective error rate of $p$ as per the common assumption in QEC[12] (Fig. 4a). Without the correction, the data qubit process fidelity linearly decreases as $p$ is increased. When the error correction is applied, errors on two and three qubits remain uncorrected, resulting in a process fidelity insensitive to $p$ up to the first order, $F(p) = 1 - 3p^2 + 2p^3$ [12]. The quadratic dependence to $p$ is a crucial property of QEC and ideally it results in an improvement of the fidelity for $p < 0.5$. We confirm this crucial property in Fig. 4b, where the measured process fidelity with the correction is plotted as the cyan curve. A polynomial fit to the data results in a coefficient of the first-order term as small as 0.01. As compared to the uncorrected encoded qubit, the corrected qubit shows improvement of the process fidelity in a range around $p < 0.45$. Although the corrected fidelities are lower than those of ideal uncorrected qubit (the black line in Fig. 4b), improvements of the coherence times and gate fidelities would ameliorate the situation. In silicon spin qubits, the intrinsic phase error is more like a quasi-static phase shift rather than a sudden phase flip. In our device, the phase shift is mainly caused by the fluctuating spins of surrounding $^{29}$Si nuclei. To demonstrate the effectiveness of our error correcting code to this type of phase error, we measure the dephasing of the encoded three-qubit state (Fig. 4c, d). As predicted from the ability to correct small phase errors in Fig. 4b, the initial slope of the fidelity decay is suppressed as compared to that of an uncorrected encoded qubit. Overall, these results show a successful implementation of three-qubit phase correcting code in silicon.

In conclusion, we have demonstrated the generation of the various three-qubit entangled states, the effective single-step resonantly driven iToffoli gate, and the fundamental properties of three-qubit quantum error correction in silicon. Extending the experiment to a larger scale would require a more flexible feedback-based correcting rotation. This would be limited by the slow spin measurement and initialization by energy-selective tunneling, which also pose a challenge to complete the error correction (or detection) before the phase coherence is completely lost. Substantial improvements should be possible by switching to the singlet-triplet readout, where high-fidelity spin measurements in a few μs [32,33], orders of magnitude shorter than the phase coherence time with dynamical decoupling[19], are routinely achieved. Along with the recent advances in scalable device design[34], electronics[35], and gate fidelities[4–6], we anticipate that it will become possible to demonstrate more sophisticated quantum error correcting codes in a large-scale silicon-based quantum processor.

**Acknowledgements**

This work was supported financially by Core Research for Evolutional Science and Technology (CREST), Japan Science and Technology Agency (JST) (JPMJCR15N2 and JPMJCR1675), JST Moonshot R&D grant no. JPMJMS2065, MEXT Quantum Leap Flagship Program (MEXT Q-LEAP) grant No. JPMXS0118069228, and JSPS KAKENHI grant Nos. 18H01819, 19K14640, and 20H00237. T.N. acknowledges support from the Murata Science Foundation Research Grant and JST PRESTO grant no. JPMJPR2017.

**Author contributions**

K.T. and A.N. fabricated the device and performed the measurements. T.N. and T.K. contributed the data acquisition and discussed the results. K.T. wrote the paper with inputs from all co-authors. S.T. supervised the project.

**Data availability**

The data that support findings in this study will be available from the Zenodo repository at (URL).

**Competing interests**

The authors declare that they have no competing interests.

**Correspondence**

Correspondence and requests for materials should be addressed to K.T. or S.T.

**Methods:**

**Quantum dot device.** The triple quantum dot device is identical to the one characterized in Ref. [22]. The device is fabricated using an isotopically natural, undoped Si/SiGe heterostructure. The Ohmic contacts are made by phosphorus ion implantation. Standard electron-beam lithography and lift-off techniques are used to fabricate the overlapping aluminum gates and the micro-magnet.

**Experimental setup.** The GHZ state tomography and the iToffoli gate characterization (Fig. 2) are performed using the experimental setup as described in Ref. [22]. In what follows, we detail the modified experimental setup used for the QEC experiments in Fig. 3 and 4. The sample is cooled down in a dry dilution refrigerator (Oxford Instruments Triton 300) to a base electron temperature of around 40 mK. The configuration of d.c. lines is the same as in the previous report[22]. Control pulses are generated by four Keysight M3201A arbitrary waveform generator (AWG) modules in a Keysight M9019A PXIe chassis (16 channels running at 500MSa/s). The plunger (P1, P2, and P3), barrier (B2 and B3), and sensor plunger gates are connected to the outputs of the AWG, each of which is filtered by a Minicircuits SBLP-39+ Bessel lowpass filter. The filtering results in a minimum pulse rise/fall time of approximately 15 ns. Microwave signals are generated by three vector microwave signal generators (two Keysight E8267D and a Rohde & Schwarz SGS100A with an SGU100A upconverter). Each microwave signal is single sideband I/Q modulated to prevent unintentional spin rotations due to microwave carrier leakage. Additionally, we use pulse modulation to further suppress the bleedthrough signal during the initialization and readout stages. Radio-frequency reflectometry is used for fast measurement of the charge sensor conductance. The right reservoir of the charge sensor quantum dot in Fig. 1b is connected to a tank circuit with an inductance of 1.2 µH and a resonance frequency of 181 MHz. The reflected signal is amplified and demodulated, then digitized using an Alazartech ATS9440 digitizer card.

**Three-spin initialization and measurement.** The three-spin initialization and measurement are performed as follows. The numbers $(n_1 n_2 n_3)$ indicate the respective number of electrons in the left, center, and right quantum dots. We collect 400 to 3,000 single-shot outcomes to obtain the measured probabilities. The labels (A)-(E) represent the gate voltage configurations depicted in Extended Data Fig. 1c.

1. Unload electrons in the left and center quantum dots by biasing gate voltages so that the ground state charge configuration is (001) (A). The duration is 100 µs.
2. Initialize $Q_1$ via spin-selective tunneling by biasing the voltages so that the charge configuration is near the (101)-(001) boundary (B). The duration is 750 µs.
3. Shuttle the electron in the left quantum dot to the center quantum dot by biasing the voltages so

that the ground charge configuration is (011) (C). No intentional gate voltage ramp is used. The typical pulse rise time is 15 ns due to the low-pass filter. We wait for 1 μs in the (011) configuration.

4. Initialize $Q_1$ via spin-selective tunneling by biasing the voltages so that the charge configuration is near the (011)-(111) boundary (D). The duration is 750 μs.
5. Initialize $Q_3$ via spin-selective tunneling by biasing the voltages so that the charge configuration is near the (110)-(111) boundary (E). The duration is 750 μs.
6. Qubit manipulation in the (111) configuration (F). The typical duration is 5 μs. There is an additional waiting time of 50 μs to reduce the effect of heating by the microwave pulses.
7. Readout $Q_1$ via spin-selective tunneling by biasing the voltages so that the charge configuration is near the (011)-(111) boundary (D). The total duration is 600 μs. The data for readout is collected for the first 200 μs. The additional waiting time of 400 μs duration facilitates the initialization of $Q_1$.
8. Perform controlled rotation between $Q_1$ and $Q_2$ to project $Q_2$ state to $Q_1$ in (111). Here, we pulse the virtual B2 gate to turn on $J_{12}$ at the charge-symmetry point. Since $Q_1$ is initialized to a spin-down state during the previous readout stage, for a $Q_2$ input state $\alpha|\uparrow\rangle + \beta|\downarrow\rangle$, the resulting $Q_1Q_2$ state is $\alpha|\uparrow\uparrow\rangle + e^{i\theta}\beta|\downarrow\downarrow\rangle$, where $e^{i\theta}$ is a phase factor that does not affect the readout. The duration is 1 μs. There is an additional waiting time of 50 μs to reduce the effect of heating by the microwave pulse.
9. Readout $Q_2$ via spin-selective tunneling of $Q_1$ by biasing the voltages so that the charge configuration is near the (011)-(111) boundary (D). The duration is 200 μs.
10. Readout $Q_3$ via spin-selective tunneling by biasing the voltages so that the charge configuration is near the (110)-(111) boundary (E). The duration is 500 μs.

**Virtual gate.** The capacitive couplings between the gates are suppressed by the virtual gate technique. We measure the capacitive couplings between the gates and construct the virtual gate as follows. The crosstalk between the exchange couplings is not taken into account. The virtual gate voltages vB2 and vB3 are used to control the exchange couplings.

$$\begin{pmatrix} vP1 \\ vB2 \\ vP2 \\ vB3 \\ vP3 \end{pmatrix} = \begin{pmatrix} 1 & 0.30 & 0.54 & 0.14 & 0.17 \\ 0 & 1 & 0 & 0 & 0 \\ 0.61 & 0.35 & 1 & 0.25 & 0.31 \\ 0 & 0 & 0 & 1 & 0 \\ 0.15 & 0.10 & 0.46 & 0.31 & 1 \end{pmatrix} \begin{pmatrix} \Delta P1 \\ \Delta B2 \\ \Delta P2 \\ \Delta B3 \\ \Delta P3 \end{pmatrix}.$$

**Single- and two-qubit gates.** The single-qubit rotations about x- and y-axes are performed by applying microwave voltage pulses resonant with the Zeeman splitting of each spin qubit. The microwave voltage results in an effective out-of-plane a.c. magnetic field by the micro-magnet, which induces electric-dipole spin resonance. The spin qubits have typical resonance frequencies of 19942.6 MHz ($Q_1$), 20372.6 MHz ($Q_2$), and 20923.2 MHz ($Q_3$). We use a shaped raised-cosine pulse with a duration of 124 (62) ns to implement a single-qubit $\pi$ ($\pi/2$) pulse. For the spectroscopy measurements in Fig. 2f, we use a Gaussian pulse (truncated at $\pm 2\sigma$). The phase rotation is virtually implemented by shifting the reference phase of I/Q modulation waveform. Wherever possible, the single-qubit gates are applied in parallel. The two-qubit CZ gate is implemented by adiabatically pulsing the exchange coupling by the barrier gates. To guarantee the adiabaticity, we use a shaped cosine pulse[4] with a duration of 50 ns to implement the CZ/2 gates, which results in a nominal peak exchange coupling of 10 MHz. During the experiments in the main text, the coupling strengths are fine-tuned to account for the conditional phase accumulation due to the residual couplings of about 0.2 MHz (Extended Data Fig. 3d-f). We set the minimum interval between pulses to 20 ns to avoid the pulse interference due to reflection.

**Three-qubit iToffoli gate.** The resonantly driven iToffoli gate consists of the three stages in Extended Data Fig. 5a. In the main text, the population transfer property of the iToffoli gate is shown. For that, we set $f_{\text{Rabi}} = J/\sqrt{3}$ ($J = J_{12} = J_{23}$) so that the off-resonant rotation in the $Q_1Q_3 = |\uparrow\downarrow\rangle/|\downarrow\uparrow\rangle$ subspaces is a $2\pi$ rotation. Furthermore, in order to obtain a correct quantum action, any unwanted phase accumulations on the three-qubit state have to be calibrated out. This can be achieved by setting an appropriate exchange pulse duration of $t_{\text{tot}} = t_{\text{dc1}} + t_{\text{MW}} + t_{\text{dc2}}$ and pulse timing of $\delta t = t_{\text{dc1}} - t_{\text{dc2}}$[30]. In theory, by setting the optimal exchange pulse duration to $t_{\text{tot}} = \pi(4 + \sqrt{3} - \sqrt{13})/J$, the conditional phases between the $Q_1Q_3 = |\uparrow\downarrow\rangle$, $|\downarrow\uparrow\rangle$, and $|\uparrow\uparrow\rangle$ subspaces can be eliminated[30]. For an exchange coupling of 4.5 MHz, it is 473 ns. In the experiment, we typically use a 460-ns-long rectangular pulse, which is shorter than the theoretical length due to the finite pulse bandwidth. The microwave pulse timing $\delta t$ is then adjusted to eliminate the conditional phase between the $Q_1Q_3 = |\downarrow\downarrow\rangle$ and the other subspaces. For the subspaces where $Q_2$ spin flip does not occur, shifting $\delta t$ does not affect the outcome. In the case where $Q_2$ flips, when $\delta t = 0$, (quasi-)static phase accumulation is fully cancelled out by the spin echo effect. The conditional phase in this case can be adjusted by varying $\delta t$ because for finite $\delta t$ the echo works only partially and there is a phase accumulation of $2\pi(f_1 - f_0)\delta t$. The remaining single-qubit phase offset is removed by a virtual single-qubit phase rotation. The phase offsets on the ancilla qubits are uncalibrated in the QEC experiments, although they can be calibrated out similarly. In Extended Data Fig. 5b, we illustrate the experimental sequence to calibrate the iToffoli gate phase accumulation. Extended Data Fig. 5c shows an example of uncalibrated iToffoli gate and Extended Data Fig. 5d shows a phase measurement after the calibration. In the QEC experiments, this calibration is performed right before the data acquisition to minimize the influence of the slow drift of the resonance frequencies.

**Readout error removal.** For each of the experiments where the readout errors are removed, we perform a reference measurement to obtain the readout fidelities. The spin-down (up) readout fidelity $F_{\downarrow i}$ ($F_{\uparrow i}$) is directly obtained by preparing a spin-down (up) state and a projective measurement of $Q_i$. Using the measured readout fidelities, we correct the raw probabilities $\boldsymbol{P}_M = (P_{\downarrow\downarrow\downarrow}, \ldots, P_{\uparrow\uparrow\uparrow})$ as $\boldsymbol{P} = (F_1 \otimes F_2 \otimes F_3)^{-1} \boldsymbol{P}_M$, where $F_i = \begin{pmatrix} F_{\downarrow i} & 1 - F_{\uparrow i} \\ 1 - F_{\downarrow i} & F_{\uparrow i} \end{pmatrix}$ and $\boldsymbol{P}$ is the corrected probabilities used for maximum likelihood estimation.

**Three-qubit quantum state tomography.** Due to the noise in the experiment, the density matrix obtained by a linear inversion is not always physical. Therefore, we use a maximum likelihood estimation to restrict the density matrix to be physical. We start from a Cholesky decomposition of a physical density matrix $\rho = T^\dagger T / \text{Tr}(T^\dagger T)$, where $T$ is a complex lower triangular matrix with real diagonal elements. $T$ has 64 real parameters $\boldsymbol{t} = (t_1, \ldots, t_{64})$ and we minimize the cost function

$$C(\boldsymbol{t}) = \sum_{\nu=1}^{64} \frac{(\langle \psi_\nu | \rho(\boldsymbol{t}) | \psi_\nu \rangle - P_\nu)^2}{2\langle \psi_\nu | \rho(\boldsymbol{t}) | \psi_\nu \rangle},$$

where $P_\nu$ is the measured probability projected at a basis $|\psi_\nu\rangle$. To determine the 64 parameters, the projection outcomes for linearly independent pre-rotations $(I, X/2, Y/2, X)^{\otimes 3}$ are used. To remove the error that could be introduced by the X pre-rotation, the projection outcomes for the X pre-rotations are calculated from the corresponding I rotation outcomes[36].

**Measurement of the iToffoli gate truth table.** To constrain all the elements of the truth table to be non-negative, we use a maximum likelihood procedure as follows. The input is a set of 64 measured probabilities $P_{ij}$ where the input is the $i$-th eigenstate and the measurement is projected at the $j$-th eigenstate. The readout errors are removed following the procedure above. We then minimize a cost function $C(P_{11}^{\text{MLE}}, \ldots, P_{88}^{\text{MLE}}) = \sum_{i,j=1}^{8} (P_{ij}^{\text{MLE}} - P_{ij})^2$ for non-negative parameters $P_{ij}^{\text{MLE}}$. We constrain $P_{ij}^{\text{MLE}}$ so that the sum of probabilities in one cycle of data acquisition is unity, i.e., $\sum_{j=1}^{8} P_{ij}^{\text{MLE}} = 1$.

**Quantum process tomography.** In the QEC experiments (Fig. 3 and 4), we perform quantum process tomography on $Q_2$ to obtain the process fidelities. The input state $|\psi\rangle$ is prepared by a spin-down initialization followed by a single-qubit rotation $R_i \in (I, X/2, Y/2, X)$. After the QEC protocol, tomographic readout of the resulting state is performed by applying pre-rotations $(I, X/2, Y/2)$ and a projective measurement. From the measured data, we calculate the outcome for X pre-rotation and remove the readout infidelities. For a quantum operation $E$ acting on a single-qubit input density matrix $\rho_{\text{in}}^k$, the density matrix of the output state can be written as follows,

$$E(\rho_{\text{in}}^k) = \sum_{m,n=1}^{4} B_m \rho_{\text{in}}^k B_n^\dagger \chi_{mn}, \tag{1}$$

where $\chi$ is the process matrix defined with respect to the Pauli operators $B = (I, \sigma_x, \sigma_y, \sigma_z)$. Linear-inversion of Eq. (1) can be performed to obtain a process matrix, however, the process matrix obtained in this way does not necessarily satisfy the physical conditions due to the noise in the experiment. As in the state tomography, we can obtain an estimate of physical process matrix by a maximum likelihood estimation. We start from a Cholesky decomposition $\chi = S^\dagger S / \text{Tr}(S^\dagger S)$ where $S$ is a lower triangular matrix with real diagonal elements. $S$ is parametrized by 16 real parameters $\boldsymbol{s} = (s_1, \cdots, s_{16})$ and we use a cost function $L(\boldsymbol{s})$ as follows,

$$L(\boldsymbol{s}) = \sum_{k,l=1}^{4} \left[ P_\downarrow^{kl} - \sum_{m,n=1}^{4} \chi_{mn} \text{Tr}\left(M_l B_m \rho_{\text{in}}^k B_n^\dagger\right) \right]^2, \qquad (2)$$

where $P_\downarrow^{kl}$ is the measured spin-down probability when an input state $\rho_{\text{in}}^k$ is prepared and an observable $M_l$ is measured. We numerically minimize the cost function to obtain the most likely estimate of physical $\chi$. Then the process fidelity relative to an identity operation is calculated as $\text{Tr}(\chi)$.

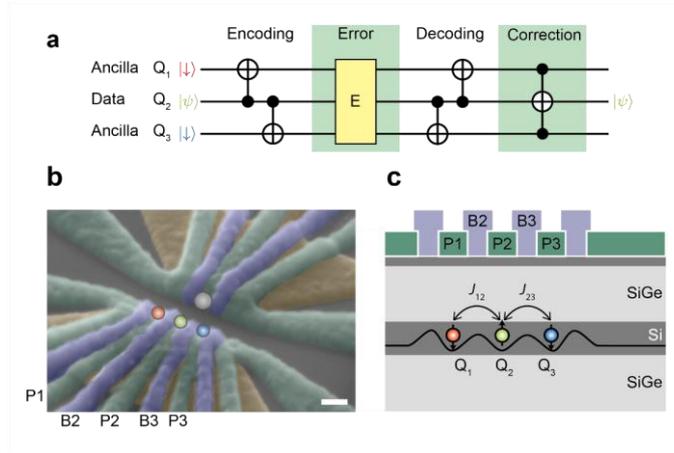

**Figure 1. Three-qubit QEC and silicon-based three-qubit device. a,** Outline of the three-qubit quantum error correcting code. Here the encoding and decoding are performed for a bit-flip error. Note that, since a phase-flip error can be converted to a bit-flip error by a single-qubit $\pi/2$ rotation, the bit-flip code is essentially equivalent to the phase-flip code. **b,** Scanning electron microscope image of the device. Scale bar, 100 nm. The screening gates (brown) are used to restrict the electric field of the plunger (green) and barrier (purple) gates. The three circles (red, green, and blue) indicate the position of the triple quantum dot array. An additional quantum dot shown as the gray circle is used as a charge sensor. The gates P1, P2, P3, B2, and B3 are connected to an arbitrary waveform generator to apply fast voltage pulses. The microwave control pulse for electric-dipole spin resonance is applied to the lower screening gate. **c,** Schematic cross-section of the device. The line in the silicon quantum well shows the schematic triple-dot confinement potential. $J_{12}$ ($J_{23}$) represents the nearest neighbor exchange coupling between $Q_1$ and $Q_2$ ($Q_2$ and $Q_3$).

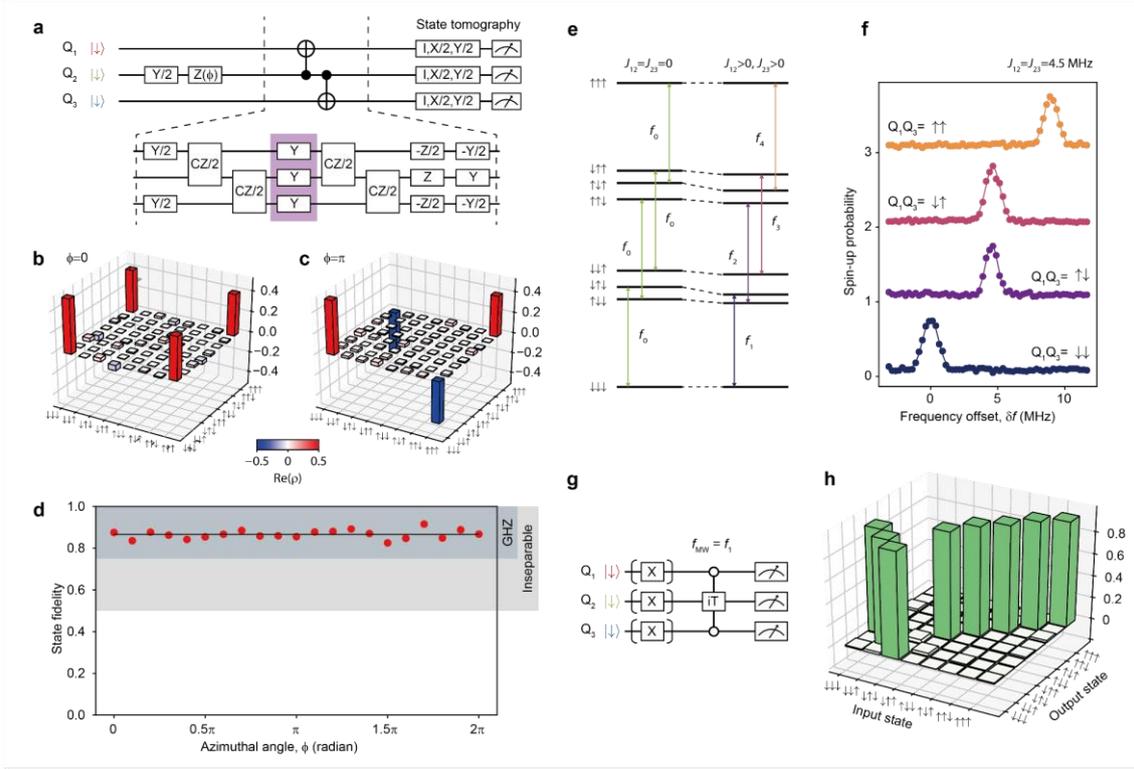

**Figure 2. Encoding of three-qubit GHZ states and resonantly driven iToffoli gate. a**, Quantum circuit to generate three-qubit GHZ-class states. X (Y, Z) represents a $\pi$ rotation about the x- (y-, z-) axis and X/2 (Y/2, Z/2) represents a $\pi/2$ rotation about the x- (y-, z-) axis. The two CNOT gates acting on the neighboring qubits are implemented by the combination of single- and two-qubit gates as shown in the bottom half. The Y pulses in the middle of the sequence (surrounded by the dashed purple line) is used to suppress the low-frequency single-qubit phase noise. **b**, **c**, Real parts of the measured density matrices of the three-qubit GHZ states ($\phi = 0$ in **b** and $\phi = \pi$ in **c**). **d**, Result of the GHZ state generation for various input states. The black solid line shows the average of GHZ state fidelities, that is 0.866. The range above the threshold value 0.75 (0.5) to distinguish the GHZ-class states from the W-class (biseparable) states is shown as the colored band. **e**, Schematic energy diagram of the three-spin state. **f**, Resonance peaks of $Q_2$ for four different control qubit states at the exchange couplings $J_{12} = J_{23} = 4.5$ MHz. Here we define $\delta f = 0$ as the resonance condition when $Q_1Q_3 = |\downarrow\downarrow\rangle$. The circles show the measured $Q_2$ spin-up probabilities for the four different control qubit configurations. The solid lines show fitting with Gaussian functions. The traces are offset by 1 from each other for clarity. **g**, Schematic sequence of the measurement of the iToffoli gate truth table. **h**, Measurement result of the iToffoli gate truth table.

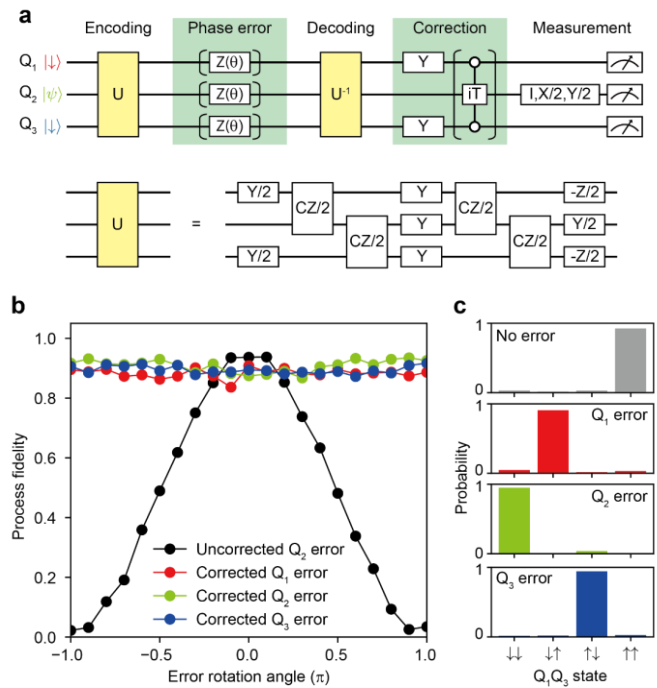

**Figure 3 One-qubit phase error correction. a**, Schematic of the quantum circuit. The operation U used for encoding and decoding is decomposed into the single- and two-qubit gates as shown in the lower half of the figure. **b**, Result of one-qubit phase error correction. In the case of uncorrected, we omit the iToffoli gate and the rest of the quantum circuit is the same as the one for the corrected case. For the ideal case without gate infidelities, the uncorrected fidelity oscillates from 0 to 1 and the corrected fidelities are always 1. **c**, Ancilla qubit measurement results. Note that due to the implementation of our correcting procedure, the resulting population of ancilla qubit state is flipped as compared to the implementation using a standard Toffoli gate[9].

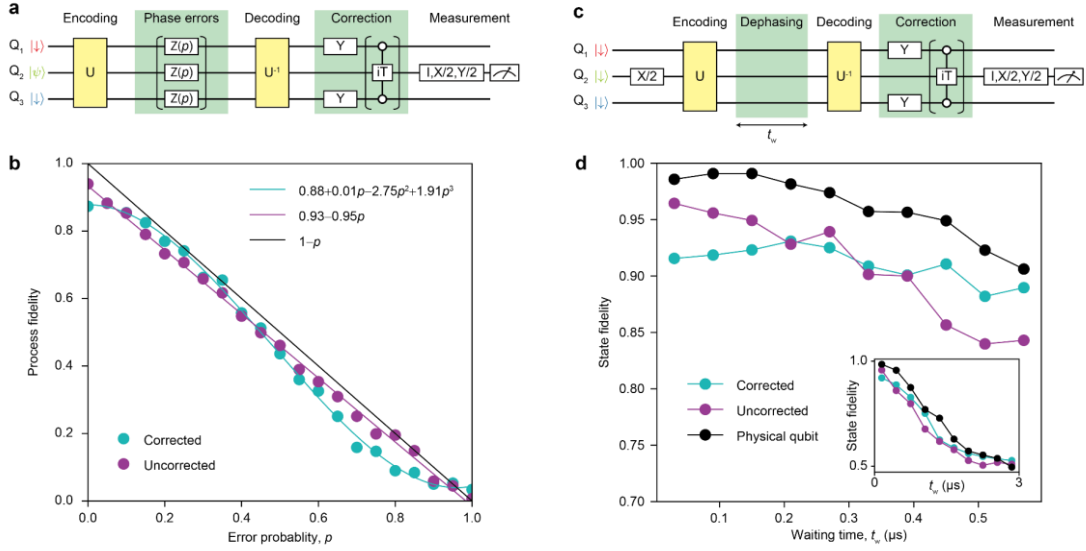

**Figure 4 Three-qubit phase error correction. a**, Schematic of the quantum circuit for three-qubit phase error correction. The phase error $Z(p)$ is a virtual phase rotation with a rotation angle of $\text{Arcsin}(\sqrt{p})$, which results in an effective error rate of $p$. We prepare the data qubit input state $|\psi\rangle$ by initialization to a spin-down state and a subsequent single-qubit rotation I, X/2, Y/2, or X. **b**, Measured process fidelities for the corrected and uncorrected cases. **c**, Schematic of the quantum circuit for three-qubit dephasing error correction. The waiting time $t_w$ is the time interval between the last single-qubit rotation in $U$ and the first single-qubit rotation in $U^{-1}$. The deviation of the purple curve from the black curve reflects the gate infidelities in the encoding and decoding. **d**, Comparison of the state fidelities of the corrected and uncorrected qubits. In the case of physical qubit, we perform a Ramsey measurement with varying a waiting time $t_w$ between the first $\pi/2$ pulse and the pre-rotation for tomographic readout. The inset figure shows the measurement for longer waiting times up to 3 μs. All states saturate to a completely mixed state with a fidelity of 0.5. The data acquisition time is the same for all traces in this figure. Each data point is obtained by averaging 3,000 experiments that are segmented into 1,000 experiments with interleaved qubit frequency calibrations.

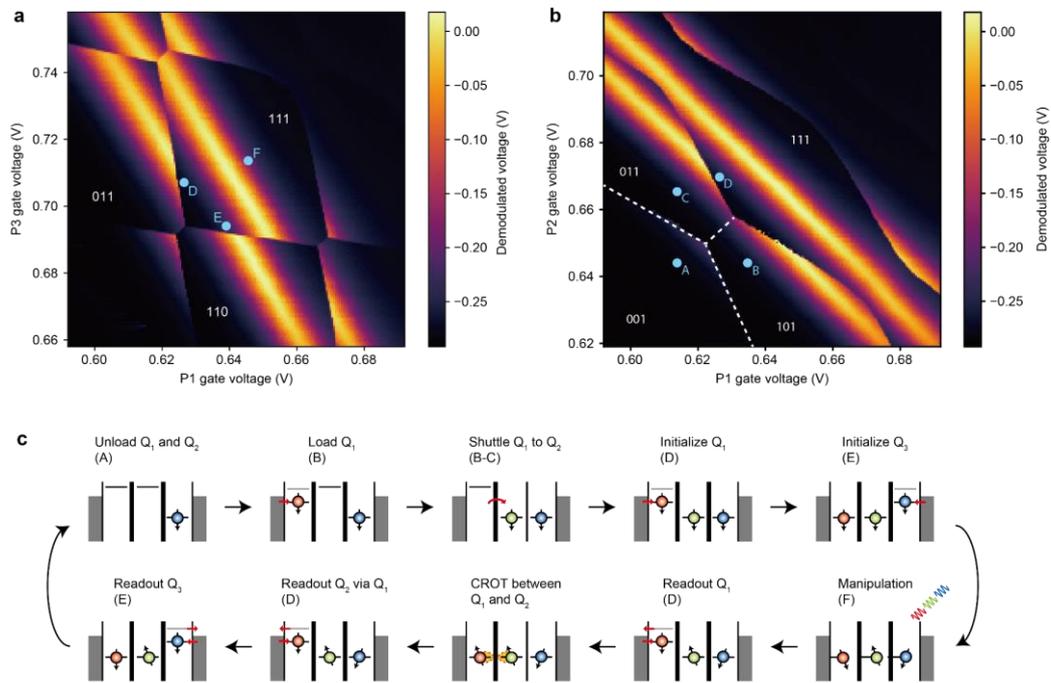

**Extended Data Figure 1. Three-spin initialization and measurement.** The numbers ($n_1 n_2 n_3$) represent the respective electron occupations in the right, center, and left quantum dot. The light blue circles with label (A-F) show the initialization, readout, and manipulation bias configurations. **a,** Charge stability diagram measured as a function of the P1 and P3 gate voltages. The variation of the background signal is due to the Coulomb oscillation of the sensor quantum dot. **b,** Charge stability diagram measured as a function of the P1 and P2 gate voltages. The white dashed lines are eye guides for the position of faint charge transition lines, which could be visible by retuning of the sensor quantum dot. **c,** schematic of the three-spin initialization and measurement.

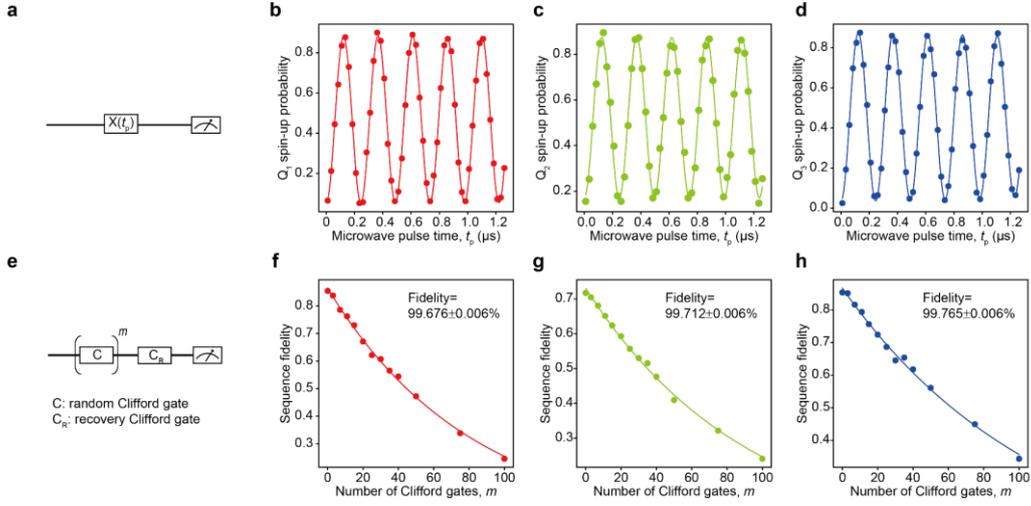

**Extended Data Figure 2. Single-qubit rotations.** All measurements are performed with all qubits initialized to spin-down and the exchange couplings turned off. **a,** Rabi oscillation measurement sequence. $t_p$ is the duration of the microwave pulse. **b-d,** Rabi oscillation measurement results. The microwave amplitude is adjusted so that the Rabi frequency is 4 MHz. **e,** Schematic sequence of the randomized benchmarking measurement. We prepare 16 randomly generated Clifford gate sequences and average the outcomes to obtain the sequence fidelities. **f-h,** Randomized benchmarking results. The implementation is the same as in, for example, Refs. [5,22]. We perform two sets of benchmarking measurements, one designed to obtain an ideal spin-up outcome and the other designed to obtain an ideal spin-down outcome, wherein both cases, the measurement is projected at a spin-up state. The sequence fidelity $F(m)$ is then defined as $F(m) = F_\uparrow(m) - F_\downarrow(m)$, where $F_\uparrow(m)$ ($F_\downarrow(m)$) is the measured sequence fidelity for the spin-up (-down) final state. Each data set is fit by an exponential decay $F(m) = Vp^m$ to extract the depolarizing parameter $p$ and visibility $V$. The fidelity shown in each figure is obtained as $1 - (1 - p)/(2 \times 1.875)$, where the factor 1.875 is the average number of primitive gates per one Clifford gate. The errors are $1\sigma$ from the mean.

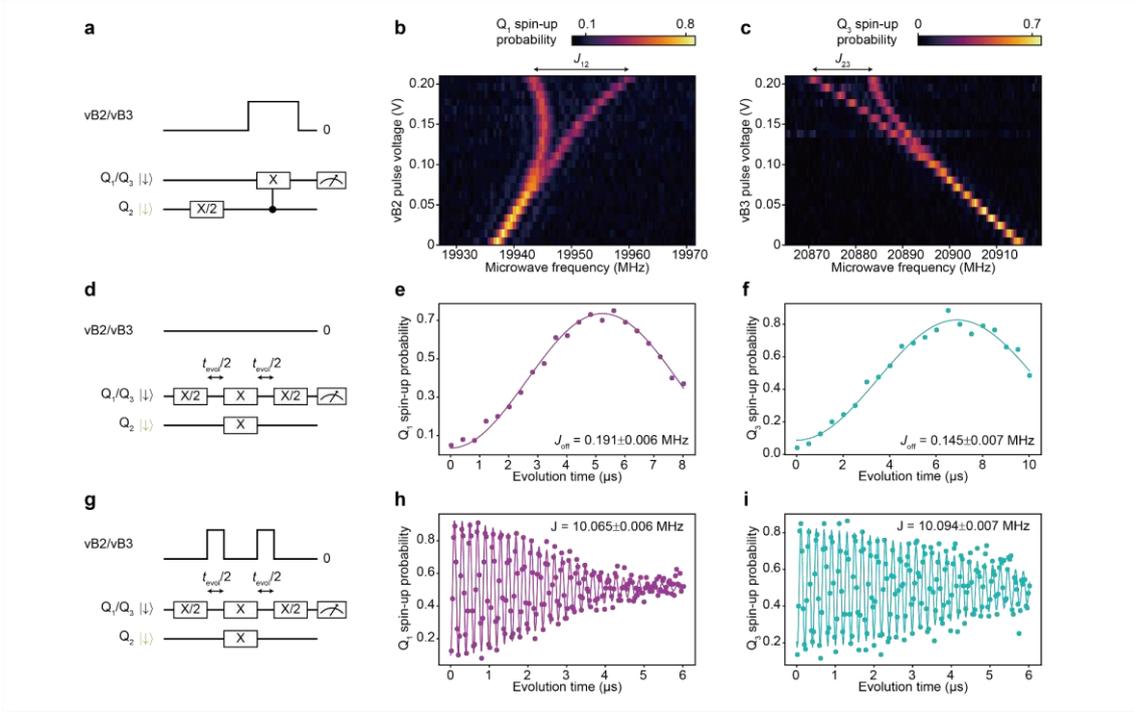

**Extended Data Figure 3. Two-qubit couplings.** All measurements are performed with all qubits initialized to spin-down. **a**, Schematic sequence of the exchange spectroscopy measurement. To narrow the resonance peaks, the microwave power for the controlled rotation is decreased by 12 dB from the values used for single-qubit rotations. $vB_i (i = 2, 3)$ represents a virtual barrier gate voltage. **b**, **c**, Results of the exchange spectroscopy measurements. In each figure, the separation of the two peaks corresponds to the exchange coupling. The background slope of the resonance frequency is due to the displacement of the quantum dot position in the micro-magnet field gradient. The frequency offset from the values in Methods is due to the decay of the persistent current in the superconducting magnet. **d**, Schematic sequence of the residual exchange coupling measurement. **e**, **f**, Results of the measurement of residual exchange couplings between neighboring qubits. Each data set is fit with a sinusoidal function $P(t_{evol}) = V\sin(\pi t_{evol} J_{off})$ to extract the residual exchange coupling $J_{off}$. $V$ is the visibility of the oscillation. The errors are $1\sigma$ from the mean. **g**, Schematic sequence of the decoupled CZ oscillation measurement. **h**, **i** Typical decoupled CZ oscillations. The solid lines show fit to a Gaussian decay. The decay times are $3.27 \pm 0.08$ μs (**g**) and $5.2 \pm 0.3$ μs (**h**). Here we adjust the virtual barrier gate voltages so that the exchange coupling is ~10 MHz. All errors are $1\sigma$ from the mean.

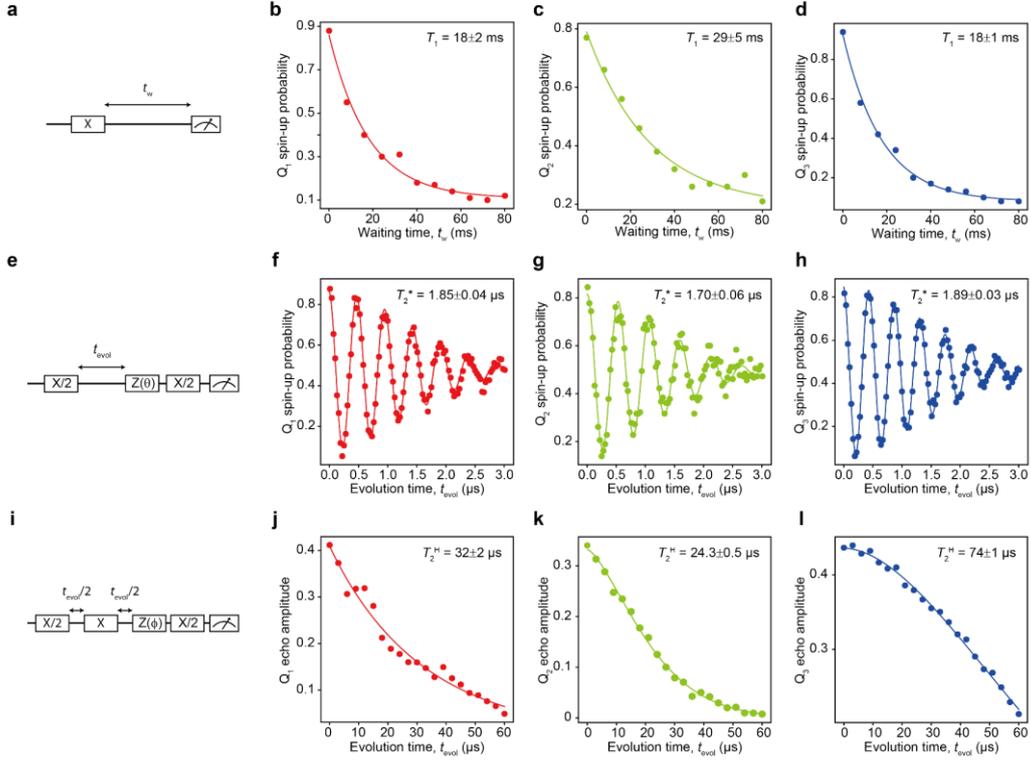

**Extended Data Figure 4. Coherence times.** All measurements are performed with all qubits initialized to spin-down and the exchange couplings turned off. All errors are $1\sigma$ from the mean. **a**, Schematic sequence of the $T_1$ measurement. The qubit state is measured after the preparation of a spin-up excited state and an idle time of $t_w$. **b-d**, $T_1$ measurement results. Each data set is fitted by an exponential decay to extract the $T_1$ relaxation time. **e**, Schematic sequence of the Ramsey interferometry. Instead of detuning the microwave frequency, we vary the phase of the second microwave pulse as $\theta = 2\pi t_{evol} \times (2 \text{ MHz})$ such that we observe an oscillation at about 2 MHz. **f-h**, Ramsey interferometry measurement results. To extract the $T_2^*$ inhomogeneous dephasing time, each data is fitted with a Gaussian decay function $P(t_{evol}) = A\exp\left(-\left(\frac{t_{evol}}{T_2^*}\right)^2\right)\cos(2\pi(\delta f)t_{evol} + \phi) + B$, where $A$ and $B$ are the constants to account for the readout fidelities, $\delta f$ is the oscillation frequency, and $\phi$ is the phase offset. The integration time is $\sim 70$ s for all traces. The larger scattering of the data points for Q$_2$ (**g**) is due to the longer pulse cycle and less averaging. **i**, Schematic sequence of the Hahn echo measurement. **j-l**, Hahn echo results. For each data set, the echo time $T_2^H$ is extracted by fitting with an exponential decay function $P(t_{evol}) = V\exp\left(-\left(\frac{t_{evol}}{T_2^H}\right)^\gamma\right)$, where $V$ is the visibility and $\gamma$ is the exponent. The exponents are $\gamma = 0.98 \pm 0.09$ (Q$_1$), $1.46 \pm 0.05$ (Q$_2$), and $1.83 \pm 0.07$ (Q$_3$).

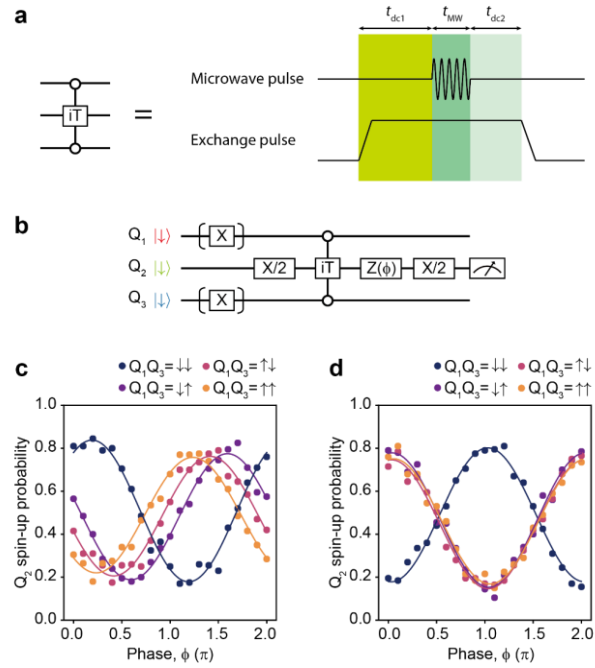

**Extended Data Figure 5. iToffoli gate phase measurement and calibration. a,** Schematic of the iToffoli gate implementation. The iToffoli gate can be realized by a combination of an exchange pulse and a microwave pulse. The exchange pulse duration ($t_{dc1} + t_{MW} + t_{dc2}$), microwave pulse duration ($t_{MW}$), and timing ($t_{dc1} - t_{dc2}$) are fine-tuned to obtain a correct phase evolution. **b,** Quantum circuit used to measure the phase accumulation during the iToffoli gate. The iToffoli gate is interleaved between two $\pi/2$ pulses to realize Ramsey-type phase detection. Only when $Q_1Q_3=|\downarrow\downarrow\rangle$, a spin flip occurs, which is detected as a $\pi$ phase shift for a correct iToffoli gate. For the other ancilla qubit configurations, the phase accumulation should be zero. **c,** Example phase measurement result before the iToffoli gate phase calibration. The resonance frequency and microwave amplitude are calibrated. **d**, Phase measurement after the calibration of both conditional and unconditional phases. In the calibration procedure, we optimize the duration of exchange pulse and the timing of microwave pulse (see Methods). We obtain correct phase evolution for all ancilla qubit configurations. The phase offsets are $(1.03 \pm 0.01)\pi$, $(0.04 \pm 0.01)\pi$, $(0.03 \pm 0.01)\pi$, and $(0.05 \pm 0.01)\pi$ for $Q_1Q_3=|\downarrow\downarrow\rangle$, $|\uparrow\downarrow\rangle$, $|\downarrow\uparrow\rangle$, and $|\uparrow\uparrow\rangle$, respectively. The errors are $1\sigma$ from the mean.

|  | $Q_1$ error | $Q_2$ error | $Q_3$ error |
|---|---|---|---|
| Encoded |  | $\alpha|+++\rangle + \beta|---\rangle$ |  |
| Error | $\alpha(\sqrt{1-p}|+++\rangle + \sqrt{p}|-++\rangle)$ $+\beta(\sqrt{1-p}|---\rangle + \sqrt{p}|+--\rangle)$ | $\alpha(\sqrt{1-p}|+++\rangle + \sqrt{p}|+-+\rangle)$ $+\beta(\sqrt{1-p}|---\rangle + \sqrt{p}|-+-\rangle)$ | $\alpha(\sqrt{1-p}|+++\rangle + \sqrt{p}|++-\rangle)$ $+\beta(\sqrt{1-p}|---\rangle + \sqrt{p}|--+\rangle)$ |
| Decoded $(Q_1Q_2Q_3)$ | $(\alpha|\downarrow\rangle + \beta|\uparrow\rangle)(\sqrt{1-p}|\downarrow\downarrow\rangle + \sqrt{p}|\uparrow\downarrow\rangle)$ | $\sqrt{1-p}(\alpha|\downarrow\rangle + \beta|\uparrow\rangle)|\downarrow\downarrow\rangle$ $+\sqrt{p}(\beta|\downarrow\rangle + \alpha|\uparrow\rangle)|\uparrow\uparrow\rangle$ | $(\alpha|\downarrow\rangle + \beta|\uparrow\rangle)(\sqrt{1-p}|\downarrow\downarrow\rangle + \sqrt{p}|\downarrow\uparrow\rangle)$ |
| Corrected $(Q_1Q_2Q_3)$ | $(\alpha|\downarrow\rangle + \beta|\uparrow\rangle)(\sqrt{1-p}|\uparrow\uparrow\rangle + \sqrt{p}|\downarrow\uparrow\rangle)$ | $(\alpha|\downarrow\rangle + \beta|\uparrow\rangle)(\sqrt{1-p}|\uparrow\uparrow\rangle + i\sqrt{p}|\downarrow\downarrow\rangle)$ | $(\alpha|\downarrow\rangle + \beta|\uparrow\rangle)(\sqrt{1-p}|\uparrow\uparrow\rangle + \sqrt{p}|\uparrow\downarrow\rangle)$ |
| Error syndrome | $|\downarrow\uparrow\rangle$ | $|\downarrow\downarrow\rangle$ | $|\uparrow\downarrow\rangle$ |

**Extended Data Table 1. Evolution of three-qubit state during QEC.** The $Q_2$ input state $\alpha|\downarrow\rangle + \beta|\uparrow\rangle$ is encoded to the three-qubit state $\alpha|+++\rangle + \beta|---\rangle$. Note that compared to the standard encoding, the coefficients $\alpha$ and $\beta$ are swapped for this state. For the decoded and corrected states, we write $Q_2$ first for the sake of brevity. When the error is a coherent phase rotation $Z(\theta)$, the error coefficient $\sqrt{p}$ ($\sqrt{1-p}$) is replaced with $\cos(\theta/2)$ ($-i\sin(\theta/2)$), while the result remains essentially equivalent.